# When Trust is Zero Sum: Automation's Threat to Epistemic Agency

*Emmie Malone[1], Saleh Afroogh*[2], Jason D'Cruz[3], Kush R. Varshney[4]*

1. Lone Star College, emmie.Siobhan.Malone@lonestar.edu
2. The University of Texas at Austin, saleh.afroogh@utexas.edu
3. The State University of New York at Albany, jdcruz@albany.edu
4. IBM Research, Thomas J. Watson Research Center, krvarshn@us.ibm.com

* Corresponding author: saleh.afroogh@utexas.edu

**Abstract**

AI researchers and ethicists have long worried about the threat that automation poses to human dignity, autonomy, and to the sense of personal value that is tied to work. Typically, proposed solutions to this problem focus on ways in which we can reduce the number of job losses which result from automation, ways to retrain those that lose their jobs, or ways to mitigate the social consequences of those job losses. However, even in cases where workers keep their jobs, their agency within them might be severely downgraded. For instance, human employees might work alongside AI but not be allowed to make decisions or not be allowed to make decisions without consulting with or coming to agreement with the AI. This is a kind of epistemic harm (which could be an injustice if it is distributed on the basis of identity prejudice). It diminishes human agency (in constraining people's ability to act independently), and it fails to recognize the workers' epistemic agency as qualified experts. Workers, in this case, aren't given the trust they are entitled to. This means that issues of human dignity remain even in cases where everyone keeps their job. Further, job retention focused solutions, such as designing an algorithm to work alongside the human employee, may only enable these harms. Here, we propose an alternative design solution, adversarial collaboration, which addresses the traditional retention problem of automation, but also addresses the larger underlying problem of epistemic harms and the distribution of trust between AI and humans in the workplace.

## 1. Introduction

Ethical discussions surrounding the introduction of artificial intelligence into the workplace are often dominated by concerns about the various costs and benefits of automation. For instance, there has already been considerable attention paid to the potential harms to human dignity that go along with humans competing with, and ultimately being replaced by, artificial intelligence systems (Stefano, 2019). Humans derive a sense of meaning from their work, and from doing it well. Likewise, they feel invested in social cohesion when they are both allowed to contribute to social progress and those contributions are recognized as valuable by that society. If artificial intelligence and robotics enter the workplace and compete for jobs with human workers, then those workers risk losing this sense of meaning and social buy-in. These concerns could be aggravated using large language models such as GPT-4 or Bard in the workplace. The possibilities for these models to replace individuals' jobs through the automation of tasks which require a person's help, resulting in loss of meaningful employment and identity associated with work are enormous. As AI content is not simply distinguishable from those created by humans, the recognition and value of human contributions may be diminished, thereby diminishing the sense of worth which individuals derive from their activities. In addition, as AI replaces human roles, the disruption may lead to a loss of social cohesion and progress. Personal growth and a sense of purpose could be hampered by the lack of incentive to develop the skills that AI can replicate. Conflicts within the workplace may also contribute to ethical dilemmas related to responsibility for AI generated content.

In addition, employment brings with it material consequences for individuals and communities that lose out on income as a result of automation. The displaced worker, their family and dependents, and their community suffer from a loss of income, potentially a loss of health insurance coverage, and downstream disinvestment in community resources (when automation affects a particular geographic region). These material consequences have the potential to result in serious secondary social harms as well. For instance, lack of access to economic opportunities may contribute to distrust of social institutions, resentment, and general social unrest (Acemoglu, 2021). These problems can be compounded when patterns of automation disproportionately affect already marginalized communities (Petersen et al., 2022). If the jobs that are automated most (or most readily) tend to be those primarily occupied by women, people of color, or those of a lower economic status, then the harms of automation are distributed in inequitable ways.

Concerns over equity and automation are not new, and ethicists and AI researchers have devoted significant attention to understanding and addressing these problems. However, ethical issues surrounding artificial intelligence in the workplace are not exhausted by concerns about automation. That is, even if workers manage to keep their jobs, certain applications of artificial intelligence alongside those workers still risk causing them significant harm and still risks perpetuating identity-prejudicial injustices. By thinking of these problems solely in terms of automation and employment, researchers risk ignoring the problems which remain even when continued employment for human workers is guaranteed, and risk endorsing design approaches which enable continued harms.

We maintain that concerns about employment do not exhaust the ethical concerns raised by artificial intelligence in the workplace, because there are additional potential 'epistemic' harms which workers can and do face as a result of the adoption of artificial intelligence. These epistemic harms concern workers' sense of dignity and meaning even when they are able to retain their jobs. A focus on job retention suggests a high-level design approach focused on human-AI collaboration. However, straightforward collaboration between humans and AI still runs the risk of perpetuating these epistemic harms. Instead, we advocate thinking about ethical issues surrounding AI in the workplace in terms of the distribution of trust (between humans and AI). The distributive schema of the workplace trust economy can be more or less equitable.[1] We propose *human-AI adversarial collaboration* as a design approach to mitigate the harms of automation (even in, but not limited to, cases where workers retain their jobs) and to recognize the epistemic agency of workers. Adversarial collaboration aims to address these concerns by avoiding the circumstances in which human-AI conflict can arise within the workplace at the design-level instead of or in addition to at the organizational level.

In section one, we describe the problem of automation in terms of a distribution of trust and expertise within the workplace. In section two, we argue that epistemic harms can persist for workers even when they retain their employment and describe various scenarios in which biases in the distribution of trust within the workplace can lead to unjust and inequitable distributions of epistemic harms. Importantly, we argue that straight-forward design approaches centered on human-AI collaboration will fail to address (and could enable) those epistemic harms. Finally, in section three, we advocate for an adversarial collaboration design approach, which better recognizes the epistemic agency of workers. We describe a few ways in which this design approach could be realized and explain how they avoid the harms already discussed.

## 2. Automation & Trust

Historically, technological innovations have played an instrumental role in human development and were considered tools in assisting humans, as agents and decision-makers, in achieving their goals (Afroogh et al., 2021). However, artificial intelligence, with its current rapid growth, has the potential to invert the historical tool-agent relation between technological production and human beings. AI systems may, in some contexts, use humans as tools to assist them in attaining their goals.

---

[1]. The concept of economy of trust refers to the distribution of trust between people and AI systems at work, describing how they share in trusting each other.

Additionally, AI-driven decision and management knowledge eliminate the need for human training to improve knowledge in some fields (Beane, 2022). Human beings might think that they do not need to perform or further develop their capacity to perform tasks which will ultimately be better performed and developed by more advanced artificial intelligence algorithms. Widespread public fatalism about the inevitability of machines to outperform humans at most tasks risks dealing a long-term blow to human creativity and ingenuity. It threatens the meaning that individuals get from pursuing self-cultivation.

The potential to aggravate these challenges is illustrated by the emergence of large language models and generative AI systems (Weidinger, et al. 2021). A risk of overdependence on their suggestions and recommendations with no critical assessment is emerging as an increasing number of AI systems gain experience in text generation or image production. It could also lead to a loss of human decision making capacity and an erosion of personal responsibility. In addition, people would be discouraged from developing their skills and exploring creative solutions, which would hamper human innovation, if they were to perceive AI as superior to humans. Social biases, which might lead to more inequality, could be amplified by the uncontrolled distribution of AI generated content. Therefore, ensuring the responsible use of AI, while enabling a sustained development of people in critical thinking, creativity and ethical choices, is essential for managing these risks.

Importantly, these harms will occur whether or not these worries are justified as long as the belief in AI's epistemic superiority is widespread. While these issues have drawn considerable attention among the public and researchers, it is important to note the ways in which issues around automation are tied with issues of expertise, agency, and trust.

The central problem of automation is that employment in a particular job is a rivalrous good. At the smallest scale, you have a human worker and an artificial intelligence competing for the same job. Within this competition, artificial intelligence is capable of outcompeting the human worker on the basis of either performance or cost-savings. Since the contest for employment in this particular role is zero-sum, the worker loses out on all of the direct and indirect benefits that come with employment. On the basis of this description, we might adopt a few strategies to ameliorate the effects of automation. One solution might be to mitigate consequences for the worker by expanding the social safety net or by offering training in some other economic sector. An alternative, but not mutually exclusive, strategy would be to design the artificial intelligence with the intent that the worker will be employed alongside the AI and collaborate with it, and thereby avoid the problem. However, and importantly, this solution will not fully address the harms associated with automation. To make clear why this is the case, it might be helpful to think about the problem of automation in broader terms.

Trust, as a disposition towards others which attributes competence and truthfulness, and which promotes reliance, is a rivalrous good. In addition to being an element of individual capital, trust is also an element of social capital (Claridge, 2020). Considered as individual capital, being trusted (especially on matters concerning one's expertise) is an important good. A systemic failure to be trusted when one is trustworthy is damaging to one's sense of dignity and achieving the trust of others allows for other goods. For instance, trust is a fundamental aspect of all communication, whether it be human-to-human or AI-to-human. There is a remarkable value and function to trust in human communities, organizations, and society. Distinguished as a type of social capital, trust can also assist local leaders, administrators, and governments in reaching their goals by enhancing the efficiency of local communities, organizations, and societies (Afroogh, 2022). Furthermore, individual and communal economic and social rivalry and competition are heavily influenced by trust's value as social capital. On the other hand, trust is a scarce good and cannot be assigned to all individuals or agents in a society. There is a direct correlation between the scarcity of trust and its value. Individuals are in competition for being trusted. In some contexts, even if there are several trustworthy agents, not all will be credited with being the trustee.

Suppose a doctor prescribes an approach that is contrary to computer programs' recommendations. The doctor's prescribes may not be followed if both recommendations are incompatible, irrespective of their compatibility with the patient's values or general competence. Similarly, an employer might choose to hire an artificial intelligence system rather than hiring a human employee. In these cases, relying on one party means not relying on the other. Ultimately, users (whether employers or patients) have to adopt one recommendation or the other where those recommendations are mutually exclusive. Insofar as trustees are interested in having their recommendations adopted by the user and adoptions are rivalrous, various schemes of adoption open themselves up to being more or less just. Situations such as these are what philosophers like John Rawls call 'the circumstances of justice' (Rawls 1971).

Granted, trust is not *always* rivalrous. There are certainly cases where one trusts several parties with respect to the same decision. Furthermore, there are cases in which one can take into account conflicting advice by synthesizing it into complex plan of action. (For example, you may decide to take risk while still hedging your bets). Nonetheless, there will be contexts where by opting to follow the recommendation of one party, you thereby reject the recommendation of the other. In many such cases, the party whose advice is jettisoned may suffer a kind of dishonor (which may or may not be warranted).

On this description, the problem of automation is one of how trust is distributed. Since adopting a particular recommendation involves trusting one party over another, the question of whether a worker will retain their influence with the end user, and thus their employment, comes down to whether their expertise is recognized by that user. However, even when the human worker retains their employment, they still may end up with a smaller share of trust than they started with as a result of the introduction of artificial intelligence into the workplace. For instance, if employers keep their human workforce but supplement it with artificial intelligence, then those AI-systems may take on a decision-making or decision-verifying role. In this case, a worker may continue to do their job, but the role that that job picks out may become increasingly circumscribed. If artificial intelligence performs at a high level, then employers may demand that workers consult with, defer to, or come to agreement with the artificial intelligence before acting. This circumscribed role fails to recognize the epistemic agency of the worker as a relevant expert who is sufficiently trustworthy in and of themselves.

## 3. Epistemic Harm & Trust Equity

We can think of this failure to recognize the complete epistemic agency as an instance of what Miranda Fricker calls an 'epistemic harm' (Fricker, 2009). This highlights the way in which the meaning and dignity that humans derive from their work isn't reducible to being able to say that they have a job or that they do work. Rather, much of the meaning and dignity that workers feel is the result of the exercise of their agency. If workers are reduced to serving at the discretion of the algorithm, then they are liable to experience alienation even if they avoid the economic consequences of a loss of income. While employers may insist that the artificial intelligence is just there to double-check their work, if correspondence between the decisions or recommendations of the worker and the algorithm is the difference-maker in whether a particular course of action is adopted, then the worker is serving in a subordinate role. Recognizing that worker's epistemic agency means trusting them as a relevant and trustworthy expert to make decisions and be held accountable for those choices. When outcomes issue from the exercise of this agency, that worker can reasonably be accountable for their success and the benefits that result from it. This is what enables a sense of meaning and dignity, values that are not necessarily realized by guaranteeing continued employment.

In addition to the epistemic harms that may face individual workers, the distribution of these epistemic harms may result in more of less equitable patterns of trust. When one fails to recognize the complete epistemic agency of another on the basis of some identity prejudice, we can think of this as an epistemic injustice (Fricker, 2009). For instance, if we assign less credence to the testimonies of women than men on the basis of an (explicit or implicit) gender prejudice, then an epistemic injustice has occurred. So-called 'testimonial injustices' of this kind can be thought of as unjust distributions of trust in cases where those testimonies are mutually exclusive because we distribute that rivalrous good according to what we, upon reflection, deem to be an unjust bias.

This kind of unjust distribution can happen in terms of employment. For instance, consider a case in which two patients decide whether to consult with one of 1) a male human doctor, 2) a female human doctor, or 3) an artificial intelligence system. In a situation where all three options are deemed sufficiently trustworthy, the decision of who each patient will trust might be rationally underdetermined all things considered. This leaves room for bias to be the difference maker in which of the three doctors is left with no patient. It has been shown that patients rate female doctors less favorably than male doctors (Wallace & Paul, 2016; Kauff et al., 2021). This means that an artificial intelligence system could be trusted more than female doctors but less than male doctors. If this is reflected in patients' adoption rates, then that artificial intelligence system will automate the roles of female doctors but not male doctors. In cases where all three are trustworthy and patients trust male human doctors and artificial intelligence systems to have the relevant medical knowledge (enough to endorse their recommendations), but not female human doctors, they fail to recognize and respect the female doctors' epistemic agency for reasons of identity prejudice.

Epistemic agency shall refer to a person's ability to acquire, interpret and make decisions on his or her own knowledge; it shall be capable of using expertise in several areas. There is epistemic damage when one does not recognize or respect a person's ethics agency. This may result in an alienation of feeling and a loss of dignity.

If automation disproportionately affects industries dominated by marginalized groups, then inequitable distributions will arise within a society even when artificial intelligence systems aren't outcompeting individuals at a biased rate within that domain. In this case, we might see differential rates of automation between male-dominated roles like medical doctors and female-dominated roles like nurses. The same patterns may result on the basis of class or race as well.

Just as the male-dominated role of doctor might be automated at a different rate than the female-dominated role of nurse, individual discretion within these roles might be given over to artificial intelligence at different rates. If doctors are allowed to use their own independent discretion when recommending diagnoses and treatment where nurses must defer to a consensus between themselves and an artificial intelligence system, then their epistemic agency is diminished in ways that we might think represent an unequitable distribution of epistemic agency. This is an important additional harm that can occur even in cases where workers retain their employment and income, and it represents an equity issue which concerns the dignity of workers as epistemic agents. Again, these same biases, but concerning race or class, might apply across or within industries.

In many collaborative contexts, trust inequity can result from unequal trust distribution caused by epistemic harm; and negatively impact the sense of agency, dignity, and meaningful participation of individuals. Inequity in trust is evident across a wide range of domains, including healthcare, legal decisions, education, workplaces, and social media.

Healthcare professionals' expertise can be undermined by reliance on AI-generated diagnoses over human specialists due to biases, resulting in misdiagnoses and inadequate treatment. Similarly, biased AI algorithms may perpetuate disparities in legal outcomes within the legal system by favoring certain demographics or discouraging others. Education and workplaces can be affected by AI recommendations undervaluing educators' expertise and alienating employees, affecting both employee satisfaction and learning. Social media algorithms that limit access to diverse perspectives can also undermine cognitive agency and understanding by exposing users to biased or polarizing content.

Furthermore, there is often a dynamics of power and group in collaborative environments. Artificial intelligence systems can reinforce existing hierarchies and social norms if they are disproportionately trusted over certain human groups. Marginalized groups may be further marginalized within these environments by limiting their opportunities to exert their expertise and agency, perpetuating inequalities. Additionally, collaboration fosters innovation and creativity, but undermining epistemic agency may devalue unique insights, hindering advancement in industries and societies, and reducing diversity in decision-making processes.

Due to these implications, epistemic harm as well as unequal trust distribution can have profound effects on a variety of collaborative settings. A collaborative partnership that harnesses the strengths of humans and robots while avoiding perpetuating injustice and bias is crucial to establishing equitable collaborations. To cultivate a fair and just relationship between humans and AI systems, it is crucial to recognize the multifaceted nature of trust inequity and its impact on agency, dignity, innovation, and societal progress.

## 4. Adversarial Collaboration

Following recent discussions regarding epistemic harm, there emerges a need for innovative approaches to mitigate such harm and ensure equitable distribution of trust in collaborative environments involving human-AI interactions. While there may be several strategies for addressing the problem of automation, by adopting design approaches which focus on harms and inequities at the level of the distribution of trust (and not employment), researchers might be able to avoid falling into the circumstances of justice altogether. One way of doing this is by designing algorithms to engage in what is called 'adversarial collaboration' with fellow trustworthy recommenders (Kahneman 2003). Human researchers use adversarial collaboration to avoid competition between defenders of opposing hypotheses. Scientists often adopt an approach where one team of researchers designs an experiment in order to disprove another team's explanation of the data and the second team responds by designing and running a responding experiment designed to disprove the first team's explanation of the data. By contrast, adversarial collaboration involves the two teams working together to design an experiment which both parties see as moving the debate forward. However, this is not mere collaboration, in which colleagues with a common point of view work together to achieve their shared ends. Instead, this approach views the adversarial stance between the two parties as essential in generating progress. We propose a similar, but distinct, model to avoid the circumstances of justice in AI-human collaboration.

One model for designing AI with human collaboration in mind might be to aim for the algorithm to perform the same task as the human agent. For instance, an algorithm may look at X-ray scans and suggest the presence of a tumor or not on the basis. This suggestion could be offered to the human doctor who makes their own recommendation, or the human doctor might merely review the recommendations made by the AI for error. This model constitutes designing for ordinary collaboration. However, since the algorithm and human perform the same task, this opens up the possibility of eventual complete automation of the human doctor's role, or subordination of the human doctor in line with the cases of epistemic harm described above. While the workplace structure may place the human doctor in the position of epistemic authority (by giving them the final say in making a diagnosis), the design of the algorithm itself does not. Our proposal is that the algorithm should in some cases, not be designed to perform the same task as the doctor. Instead of establishing a straightforward partnership between the AI and the human, we could design the algorithm to occupy an adversarial relationship within their collaboration. The task of the AI, on this model, is not to offer an alternative recommendation, but to scrutinize the basis for the human's decision. This might be to identify counterevidence, to offer alternative explanations of evidence in favor of the human's decision, or to mount the best defense of the alternative position. All of these approaches would constitute models of adversarial human-AI collaboration. The human agent could then engage in an iterative process with the AI for a number of cycles before arriving at a final decision.[2]

Our proposal resembles Kahneman's model in some ways and departs in others. As the name suggests, they share in common that the parties are aligned in their goal (arriving at an optimal decision) but are oriented toward critically scrutinizing each other's views. However, Kahneman models human-human collaboration in science and our proposal is designed to model human-AI collaboration. This introduces an important distinction. Human-human collaboration involves multiple agents, each with their own epistemic agency deserving of recognition. This puts both parties in an equal position of epistemic authority. Our model involves human-AI collaboration and is designed specifically to protect the epistemic agency of the human agent. This problem is solved by creating an asymmetry of epistemic authority which places the human in the primary position. The human is ultimately making the decision and the algorithm isn't designed to compete in this role, but to interrogate and, thus, sharpen the recommendations of the human agent.

This situation avoids the circumstances of justice because users aren't forced to make a mutually exclusive choice between trusting the algorithm or the human. The problem of trust equity only arises under the circumstances of justice, where humans and artificial intelligence systems compete for scarce user attention. In a model of adversarial collaboration between human and artificial intelligence recommenders, both parties work together to achieve the user's adoption of their joint recommendation. By contrast, less adversarial forms of collaboration between humans and AI (in which, for example, an AI might offer a doctor a recommendation that that doctor could take or leave), we run the risk of making the doctor redundant should the AI outpace the doctor's performance. That is, there is no safeguard against collaborative AI and their human users entering into the circumstances of justice. Meanwhile, the adversarially collaborative AI's contribution to the recommendation is necessarily subordinate to the human who makes both the initial and final recommendation to the end user.

---

[2]. An alternative approach to adversarially collaborative design might have human agents play the antagonistic role with AI systems are the lead recommender. This model has already been proposed for reasons unrelated to trust equity (Attenberg et al., 2015).

Because this model avoids the situation where AI systems outcompete some or all workers for jobs as expert recommenders, it avoids the potential for total automation and the potential harms that go with it (in terms of human dignity and/or income). Likewise, because the human agent still plays an essential (and primary) role in providing end users with recommendations, adversarially collaborative design works to recognize the epistemic agency of expert workers and the epistemic injustices associated with a failure to do so. Recognizing workers' epistemic agency does not need to involve blind trust. Indeed, the algorithm taking the human's reasons and recommendations seriously, and providing some scrutiny, is, in a way, a recognition of their epistemic agency. In this way, artificial intelligence can contribute to human problem-solving. The moral demands of epistemic agency merely require that human be challenged by alternative evidence and explanations (which can be provided by artificial intelligence), to incorporate that feedback, and take responsibility for their own resulting conclusion.

Importantly, such a model (but using humans) is already proposed as the next crucial scientific reform needed to address the persistent weaknesses in social scientific norms, and similar but distinct strategies for AI-human collaboration have shown some success in medical imagining (Clark, 2023; Leibig et al., 2022). In this case, the artificial intelligence system developed according to an adversarially collaborative design approach plays the role of a 'devil's advocate'. Given these risk-mitigating effects, we might expect an increase in the trustworthiness of the resulting recommendation. This, in turn, could be reflected in users' willingness to trust the recommendation. In addition to avoiding the circumstances of justice, the benefits of the improved recommendation which results from the adversarial collaboration could produce a Pareto efficiency, raising the trustworthiness of the AI and human recommender while better serving the end user. The enhanced trustworthiness that comes with better risk management strategies could lead to wider user adoption and enhanced public opinion of AI solutions, a problem which already attracts significant attention (Akkara & Kuriakose, 2020; Chen & Wen, 2021; Jackson & Panteli, 2021; Spiegelhalter, 2020; Tschopp, 2019). This is to say that adversarial collaboration, as a high-level design approach may address concerns about automation and epistemic harms to workers (and thus avoid potentially unequitable distributions of those harms), but also work to counter the problem of trust in AI.

While epistemic harms to workers can be serious and weighty, they are not the only value that should be considered in most cases. While adversarial collaboration addresses the issue of epistemic harms, it may not be appropriate as a general design strategy in all circumstances, especially when epistemic harms might be outweighed by considerations of accuracy or end-user autonomy. For instance, in the example above, high stakes decisions about health (such as determining whether a tumor is cancerous), what matters most is that the right answer is arrived at, and that the patient is able to make informed autonomous decisions about how to proceed. These circumstances might suggest a more straight-forward collaboration model. However, where decisions are relatively low stakes, epistemic harms may become a sufficiently salient moral feature when designing an algorithm for human-AI collaboration. Take, for example, a radio DJ who must determine which set list is most likely to get listeners dancing. A well-trained AI could reasonably outperform the human at this task in all cases, but the costs associated with a slightly suboptimal dance music playlist do not warrant the associated harms to worker agency. Likewise, adversarial human-AI collaboration may be appropriate in situations where human agents have special access to sources of information that an AI will not. For example, a human doctor or nurse will know things about a patient's values and preferences as well as their ability to follow through on treatment plans. While the AI is not making an 'error' in calculation in these cases, it comes to a worse judgment about how to proceed. Deference to the algorithm (as a result of automation or subordination of the human agent) in such a case is an insult to the expertise and agency of the healthcare worker and an injury to the patient as an end-user. This is just to say that both straight-forward and adversarial models may be optimific under different circumstances, and that a one-size-fits-all approach to collaborative design may not be appropriate.

Likewise, there are contexts in which epistemic harms are more salient than others. This is especially true with respect to workers who are of lower status or are less powerful. The subordination of a nurse's expertise and agency in the workplace is not counteracted by significant social status as an expert the way that a surgeon's might be. In the same way, epistemic harms will matter more in cases where decisions have more bearing on workers' own work arrangements than on end users. For example, consider decisions about workplace shift scheduling or protocols that are determined by algorithm to maximize efficiency. The benefits of this gain in efficiency are often secondary and minimal to the end user. However, the agency and autonomy that worker's exercise in organizing their own workplace and workflow are important in recognizing their epistemic authority. These factors should be considered in determining whether adversarial collaboration is the optimal design paradigm for a given situation and task.

## 5. Conclusion and Future Directions

We have identified a limitation of thinking about the problems of automation solely in the context of employment. Namely, that doing so will fail to capture the ways in which workers who retain their employment may still be harmed as epistemic agents when their discretion is diminished as they are required to defer decision-making authority to artificial intelligence systems. Like the direct and indirect harms of automation, the harms of diminished epistemic agency in the workplace may be distributed in more or less equitable ways. By approaching the problem of automation from a higher level, in which the issue is cast as one of how trust is distributed within an ecosystem of workers, we can attend to the problems of employment, income, and epistemic agency simultaneously. This conception recommends adopting design approaches which center workers' epistemic agency, and are sensitive to the ways in which trust is distributed.

Ultimately, we suggest that designing AI systems with the goal of adversarial collaboration in mind better serves this vision than merely designing those systems with simple non-adversarial collaboration in mind. This is because adversarial collaboration recognizes the epistemic agency of expert workers and builds not just a continued role, but continued agency, into the desired use plan for artificial intelligence. Importantly, this approach avoids the pitfalls associated with a narrow "focus on application that automate jobs…" while still harnessing the power of artificial intelligence to make human decision-making and industry more reliable, more trustworthy, and more efficient (Acemoglu, 2021). As a result, adversarial collaboration represents a design strategy for "AI for good" (Acemoglu, 2021).

While we have identified two areas of concern which can arise within the ecosystem of trust in which artificial intelligence systems compete (employment and epistemic harms), more work should be done to identify further areas of concern and to address complications that emerge as a result of competition for users' trust. In addition, more work should be done to identify domains in which these harms are already occurring and where they might be occurring at inequitable rates on the basis of identity-prejudicial biases. Finally, adversarial collaboration may come with its own unique set of obligations to end users. We expect that there will be cases in which the decision maker has an obligation to divulge the counter-evidence or alternative explanations offered by the adversarial AI. There are likely a variety of factors which might make disclosures of this kind more pressing. For instance, certain medical decisions with high stakes for patient health outcomes may require more transparency about the nature of the adversarial collaboration which lead to a particular diagnosis or treatment plan where low-stakes decision making may not. Further work should be done to identify the factors which might influence how transparency is achieved for end users within an adversarially collaborative framework.

**Funding:** This research is funded by the SUNY-IBM AI Research Alliance under grant number AI2102.

**Conflict of Interest:** The authors declare that the research was conducted in the absence of any commercial or financial relationships that could be construed as a potential conflict of interest.